\documentstyle[sprocl,epsfig]{article} 

\def\ra{\rightarrow}

\newcommand{\nc}{\newcommand}
\nc{\be}{\begin{equation}}
\nc{\ee}{\end{equation}}
\nc{\bea}{\begin{eqnarray}}
\nc{\eea}{\end{eqnarray}}
\nc{\beas}{\begin{eqnarray*}}
\nc{\eeas}{\end{eqnarray*}}
\nc{\noi}{\noindent}
\nc{\sD}{\not \! \! D}
\nc{\s}[1]{\not \! #1}
\nc{\non}{\nonumber}
\nc{\bb}{\bibitem}
\nc{\lf}{\left}
\nc{\ri}{\right}
\nc{\mb}[1]{\makebox[#1]{}}
\nc{\pa}{\partial}
\nc{\sA}{\not \! \! A}
\nc{\newsec}[1]{\section{#1}\mb{0.5cm}}
\nc{\h}{\frac{1}{2}}
\nc{\la}{\leftarrow}
\nc{\ep}{$e^+e^-\ra\pi^+\pi^-\;$}
\nc{\emuon}{$e^+e^-\ra\mu^+\mu^-\;$}
\nc{\epp}{$e^+e^-\ra\pi^+\pi^0\pi^-\;$}
\nc{\elec}{$e^+e^-\ra\gamma^*\ra e^+e^-\;$}
\def\mathunderaccent#1{\let\theaccent#1\mathpalette\putaccentunder}
\def\putaccentunder#1#2{\oalign{$#1#2$\crcr\hidewidth
\vbox to.2ex{\hbox{$#1\theaccent{}$}\vss}\hidewidth}}

\nc{\ti}{\mathunderaccent\tilde}
\nc{\M}{{\cal M}}
\nc{\rw}{$\rho\!-\!\omega\;$}
 
\def\hhha{\rule[-3.mm]{0.mm}{7.mm}}
\def\hhhb{\rule[-3.mm]{0.mm}{12.mm}}

\begin{document} 


\title{NEW RESULTS IN VECTOR MESON DOMINANCE AND RHO MESON PHYSICS} 
 
\author{ A.G.\  WILLIAMS } 
 
\address{Special Research Centre for the Subatomic Structure of Matter\\
 and Department of Physics and Mathematical Physics,\\
         University of Adelaide, Australia 5005 \\
         {\rm e-mail: awilliam@physics.adelaide.edu.au}
} 
 
 
\maketitle\abstracts{ 
Recent results by Benayoun {\it et al.}
comparing the predictions of a range of existing models based
on the Vector Meson Dominance hypothesis are summarized and discussed.
These are compared with data on
$e^+e^-\rightarrow \pi^+\pi^-$ and $e^+e^-\rightarrow \mu^+\mu^-$
cross-sections and the phase and near-threshold behavior of the timelike
pion form factor, with the aim of determining which (if any) of these
models is capable of providing an accurate representation of the full
range of experimental data.  We find that, of the models considered,
only Hidden Local Symmetry (HLS) model proposed by Bando {\it et al.}
is able to  consistently account for all
information, provided one allows its  parameter $a$ to vary from
the usual value of 2 to 2.4.  Our fit with the HLS model gives
a point--like coupling $\gamma \pi^+ \pi^-$
of magnitude $\simeq -e/6$, while the common formulation of
VMD excludes such a term. The resulting values
for the $\rho$ mass and $\pi^+\pi^-$ and $e^+e^-$
partial widths as well as the branching ratio for the decay
$\omega \rightarrow \pi^+\pi^-$ obtained within the context of this model are
consistent with previous results.} 
 
\section{Introduction.}

A recent detailed study \cite{benayounetal}
by Benayoun {\it et al.}\ of a variety of models of vector meson dominance and
associated descriptions of the $\rho^0$
meson is reviewed and summarized.
The aim is to find an optimum modelling able to account
 most precisely  for  the
known features of the physics involving this meson.
It is important to emphasise that our philosophy is to look for
the simplest models as these are the most useful in
application to other systems, due to their ease of implementation.
Naturally, such models should, as much as possible, respect
basic general principles such as gauge invariance and unitarity.
One must keep in mind that any parameters quoted for a given
model are relevant only to that model.
Indeed, a study of the model-dependence of resonance parameters is one
of the principal goals of this work.

We study the strong
interaction corrections to one-photon mediated processes
in the low energy region where QCD is non-perturbative.
To do this we shall look at two related processes, \ep and \emuon.
The effect of the
strong interaction is obvious in the first reaction and provides a
large enhancement to the
production of pions in the vector meson resonance region
\cite{orsay,benaksas,barkov,kurdadze}.
This enhancement,
relative to what would be expected for a structureless, pointlike
pion, is reflected in the deviation of the pion form factor, $F_\pi (q^2)$,
from $1$, and is primarily associated with the $\rho$ meson (where
$q_\mu$ is the four momentum of the virtual photon).
This form factor is successfully modelled in the intermediate energy
region using the vector meson dominance (VMD) model
\cite{review}. VMD assumes that the photon interacts with
physical hadrons through vector mesons and it is these
mesons that give rise to the enhancement, through their resonant
(possessing a complex pole) propagators of the form
\be
D_{\mu\nu}(q^2)=\frac{-g_{\mu\nu}}{q^2-{m}_V^2+i{m}_V\Gamma_V
(q^2)},
\ee
where ${m}_V$ and $\Gamma_V$ are the (real valued) mass and
the momentum-dependent width.
(Here we have included only that part of the propagator
which survives when coupled to conserved currents.)

Traditionally, VMD assumes that {\em all} photon--hadron coupling
is mediated by vector mesons.
However, from an empirical point of view,
one has the freedom, motivated by Chiral Perturbation Theory (ChPT)
to include other contributions to such interactions.  The values
extracted from $F_\pi(q^2)$ for the
$\rho$ mass, $m_\rho$, and width, $\Gamma_\rho$, are
model-dependent and in quoting values
for them, the model used should be clearly stated.  

We thus turn our attention to \emuon. In modelling the strong
interaction correction to the photon propagator, VMD assumes that the
strong interaction contribution is saturated by the
spectrum of vector meson resonances \cite{W}.  Therefore,
in principle, we can extract information on the vector meson parameters
(independently of the \ep fit) without having to worry
about non-resonant processes. However, as the vector mesons enter
in \emuon with an extra
factor of $\alpha$ compared with \ep, their contributions are considerably
suppressed,
making their extraction difficult.  For this reason, we shall perform a {\em
simultaneous} fit to both sets of data, in order to impose the best possible
constraint on the vector meson parameters, and see if existing muon data are
already precise enough in order to constrain the $\rho^0$ parametrisation.

Another way to constrain the  descriptions of the
$\rho^0$ meson is to compare the strong interaction
$\pi \pi$ phase obtained using the various VMD parametrisations determined
in fitting \ep with the corresponding phase \cite{petersen} obtained
using  $\pi \pi$ scattering
{data} and the general principles
of quantum field theory, as well as the near--threshold predictions of
ChPT. This happens to be more fruitful and conclusive
in showing how VMD should be dealt with in order to reach an agreement
with a large set of data and with the basic principles of quantum field
theory.

\section{Vector meson models}\label{models}

We shall now provide a description of the various
models we will use to fit
the data for both \ep and \emuon. The cross-section for \ep
is given by (neglecting the electron mass)
\be
\sigma=\frac{\pi \alpha^2}{3}\frac{(q^2-4m_\pi^2)^{3/2}}{(q^2)^{5/2}}
|F_\pi(q^2)|^2,
\label{tempex}
\ee
where the form factor, $F_\pi(q^2)$ is determined by the specific model.
Similarly, $F_\mu(q^2)$ is defined to be the form factor
for the muon, and the cross-section for \emuon is given by,
\be
\sigma=
\frac{4\pi\alpha^2}{3q^2}\sqrt{1-\frac{4m_\mu^2}{q^2}}
\left(1+\frac{2m_\mu^2}{q^2}\right)|F_\mu(q^2)|^2.
\label{cross}
\ee
It is worth noting that these standard definitions of $F_\pi(q^2)$ and
$F_\mu(q^2)$ contain all non-perturbative effects, including for example
the photon vacuum polarisation, since Eqs.~(\ref{tempex}) and (\ref{cross})
are written assuming a perturbative photon propagator.

We shall use VMD, the Hidden Local Symmetry (HLS) model \cite{bando} and
what we will refer to as the WCCZW model (a phenomenological modification
of the general framework of Ref.~\cite{WCCWZ}, based on work
by Birse\cite{birse}), as well as modifications of these models,
which cover or underlie a large class of effective Lagrangians
describing the interactions of photons, leptons and pseudoscalar and vector
mesons. Birse \cite{birse} has shown that typical effective theories
involving vector mesons based on a Lagrangian approach, such as
``massive Yang-Mills" and ``hidden gauge" (i.e., HLS-type) are equivalent.
We therefore
consider our following examination to be reasonably comprehensive.
Numerical approaches to the strong interaction in the vector meson
energy region \cite{num}, which are not based
on an effective Lagrangian involving meson degrees of freedom
do not yet possess the required calculational accuracy for our task.

\subsection{VMD}
The simplest model is VMD itself. As has been discussed
in detail elsewhere
\cite{review,OWBK,OPTW2} VMD has two equivalent formulations, which we
shall call VMD1 and VMD2.
The VMD1 model has a momentum-dependent coupling between the photon and the
vector mesons and a direct coupling of the photon to the hadronic final
state. The resulting form factor is (to leading order in isospin
violation and $\alpha = e^2/4\pi$):
\bea\non
F_\pi^{\rm VMD1}(q^2)&=&1-g^{\rm VMD1}_{\rho\gamma}(q^2)
\frac{g_{\rho\pi\pi}}{[q^2-m^2_\rho+im_\rho
\Gamma_\rho(q^2)]}\\
&&-g^{\rm VMD1}_{\omega\gamma}(q^2)
\frac{1}{[q^2-m^2_\omega+im_\omega
\Gamma_\omega(q^2)]}Ae^{i\phi_1}.
\label{ff1}
\eea

The $\omega$ enters into the isospin 1 \ep interaction with an attenuation
factor specified by
the pure real $A$ and the Orsay phase, $\phi$ \cite{MOW}; these
can be extracted from experiment\footnote{Note in Ref.~\cite{MOW}
that $A$ and $\phi$ were defined through the S matrix pole positions
(equivalent to Eq.~(\ref{ff2}) with {\em constant} widths).
In our fit procedure, $A$  is connected with the
width $\Gamma(\omega \rightarrow \pi^+ \pi^-)$ (see Ref.\ \cite{benayoun}).}
For VMD2 we have,
\be
F_\pi^{\rm VMD2}(q^2)=-g^{\rm VMD2}_{\rho\gamma}
\frac{g_{\rho\pi\pi}}{[q^2-m^2_\rho+im_\rho
\Gamma_\rho(q^2)]}-g^{\rm VMD2}_{\omega\gamma}\frac{1}
{[q^2-m^2_\omega+im_\omega
\Gamma_\omega(q^2)]}Ae^{i\phi_2}.
\label{ff2}
\ee
The form factor for the muon, however, in both representations
(i=1, 2) is given by
\be
F_\mu^{\rm VMDi}=1+\sum_V e^2 [g_{V\gamma}^{\rm VMDi}(q^2)]^2
\frac{1}{q^2-{m}_V^2+i{m}_V\Gamma_V(q^2)}\frac{1}{q^2} ,
\label{muonff}
\ee
which is consistent with previous expressions  for the
$\phi$-meson \cite{orsay,PR}.
In higher order ({\it i.e.}, in all but the minimal VMD picture)
there will also be contributions from non--resonant processes (such
as two--pion loops), but these are expected to be small near resonance and
the non--resonant background is, in any case, fitted in extractions of
resonance parameters from the experimental data.
The photon--meson coupling,
$eg_{V\gamma}$ is fixed in VMD2 by \cite{PR} \be \Gamma_{V\ra
e^+e^-}=\frac{4\pi\alpha^2}{3{m}_V^3}g^2_{V\gamma}.
\label{coupdef}\ee
The (dimensionless) universality coupling, $g_V$, is then defined by
\be
g_{V\gamma}^{\rm VMD2}={m}_V^2/g_V\label{vmd2g}
\ee for VMD2 \cite{OWBK,MOW}.
This coupling (and universality) has been most closely studied
for the $\rho$ meson. A gauge-like argument \cite{review,Sak}
suggests that the $\rho$ couples to all
hadrons with the same strength  $g_V$ (universality)
\cite{OWBK}. However, experimentally, universality is observed to be not
quite exact \cite{klingl},
so we introduce the quantity $\epsilon$ (to be fitted)
through
\be
g_{\rho\gamma}^{\rm VMD2}=\frac{{m}_{\rho}^2}{g_{\rho\pi \pi}} (1 +\epsilon)
\label{epsy}
\ee
where $g_{\rho\pi \pi}$ and $\epsilon$ are to be extracted from
the fit to \ep.
For VMD1, it can be seen that the photon-meson coupling results from
replacing the mass term in Eq.~(\ref{vmd2g}) by $q^2$ (see
Ref.\cite{review,Sak}).
In this case Eq (\ref{epsy}) should be replaced by
\be
g_{V\gamma}^{\rm VMD1}=\frac{q^2}{g_{\rho\pi \pi}} (1 +\epsilon)
\label{1vgam}~~.
\ee

One can
easily see that in VMD1 the hadronic correction to the photon propagator
goes like $q^4$ and so maintains the photon pole at $q^2=0$. For VMD2,
it is not obvious \cite{Sak_orig} that gauge invariance
is maintained until one considers the inclusion of
a bare photon mass term in the VMD Lagrangian that
exactly cancels the hadronic correction at $q^2=0$.
This argument \cite{KLZ67} assumes a $\rho-\gamma$
coupling of the form $em_\rho^2/g_{\rho}$ and a
bare photon mass ($e^2m_\rho^2
/g_{\rho}^2$) (the calculation is presented in detail in Ref.~\cite{review}).
In the presence of a finite $\epsilon$, as in Eq.~(\ref{epsy}),
gauge invariance is similarly preserved by including a photon mass
term ($e^2m_\rho^2(1+\epsilon)^2/g_{\rho}^2$) leading to a massless
photon as expected \cite{craig}.

The presence of a finite $\epsilon$
does affect the charge normalisation condition
$F_\pi(0)=1$, but this is merely an artifact of
the simple $\rho$ propagators we are considering. A
more sophisticated  version, such as used by Gounaris and Sakurai \cite{GS}
which fully accounts for below threshold behaviour,
maintains $F_\pi(0)=1$ in the presence of $\epsilon\neq0$.
One could, alternatively, include an $s$ dependence to $\epsilon$ such
that $\epsilon(0)=0$.
In any case, the phenomenological significance for
the physical $\rho$ (for the data region we are fitting) is
negligible, we achieve an excellent fit to the data with the simple
form we use but are not advocating its use outside this fitting region,
namely above the two-pion threshold.

The choice of the detailed form of the momentum-dependent
width, $\Gamma_V(q^2)$, allows one certain amount of
freedom. As the complex poles of the amplitude are field-choice
and process-independent properties of the S-matrix  \cite{poles1}
one could, on the one hand, expand the propagator as a Laurent
series in which the non-pole terms go into the background
\cite{bernicha}. Alternatively, one could use the $l-$wave
momentum-dependent width to account for the branch point structure of
the propagator above threshold ($q^2=4m_\pi^2$) \cite{benayoun,GS,PRoos}.
This form for the momentum dependent width arises naturally from the
dressing of the $\rho$ propagator in an appropriate
Lagrangian based model (see Sec.\ 5.1 of
Klingl {\it et al.} for a detailed treatment \cite{klingl}) and, for such
models, is given by
\be
\Gamma_\rho(q^2)=\Gamma_\rho\left[\frac{p_{\pi}(q^2)}{p_{\pi}(m_{\rho}^2)}
\right]^3\left[\frac{m_\rho^2}{q^2}\right]^{\lambda/2},
\label{mesnwid}
\ee
introducing the fitting parameters $\Gamma_{\rho}$
(the width of the $\rho^0$ meson at $q^2=m_{\rho}^2$) and $\lambda$, which
generalises the usual $l$--wave expression \cite{benayoun}
to model the fall--off of the $\rho$ mass distribution;
the usual case ({\it i.e.} $\lambda=1$) is associated
with a $\rho$ coupling to pions of the form
$g_{\rho\pi\pi}\rho^{\mu} (\pi^+ \partial_{\mu}
\pi^- - \pi^- \partial_{\mu} \pi^+)$, with $g_{\rho\pi\pi}$
independent of $q^2$,
as shown in Ref. \cite{klingl}. Note that the parameters $\lambda$,
$m_{\rho}$ and $\Gamma_{\rho}$ are model-dependent (as we see in the tables of
results, different models yield different values for these parameters).
Note that in Eq.~(\ref{mesnwid})
we have defined the pion momentum in the centre of mass system
\be
p_{\pi}(q^2)=\frac{1}{2} \sqrt{q^2-4m_{\pi}^2}.
\label{ppi}
\ee

Before closing this section, let us remark that the pion form
factor associated with VMD1 (Eq. (\ref{ff1})) fulfills
automatically the condition
$F_{\pi}(0)=1$ whatever the value of the universality violating
parameter $\epsilon$ (see Eq. (\ref{1vgam})). This is not the case for
the pion form factor associated with VMD2 (Eq. (\ref{ff2})), as  can
be seen from Eqs. (\ref{ff2}) and (\ref{epsy}). Here,  we will concern
ourselves exclusively with fitting data in the above threshold region
and simply note it is a relatively straightforward matter to generalise
the VMD models considered to satisfy this condition.  Detailed
considerations of this issue are left for future work.

While of course VMD1 and VMD2 are equivalent in the limit of
exact universality if one keeps {\it all} diagrams, ({\it i.e.},
if one works to infinite order in perturbation theory), in
any practical calculation one cannot do that and so, in practice,
these two expressions of VMD can give different predictions in
general, even if exact universality is imposed. Moreover, if one releases
(as we do) this last constraint, equivalence of VMD1 and VMD2
is not guaranted, even in principle.

\subsection{The Hidden Local Symmetry Model}
The HLS model \cite{review,bando} introduces a parameter $a$
for the $\rho$ meson within a dynamical symmetry breaking
model framework. This $a$ relates the
constant $g_{\rho\pi\pi}$ to the universality coupling
$g_\rho$ via
\be
g_{\rho\pi\pi}=\frac{a g_\rho}{2}.
\ee
The resulting form factor for the pion is
\be
F_\pi(q^2)=-\frac{a}{2} + 1
- g_{\rho\gamma}\frac{g_{\rho\pi\pi}}
{(q^2-m_\rho^2+im_\rho\Gamma_\rho(q^2))} - g_{\omega\gamma}\frac{1}
{q^2-m^2_\omega+im_\omega
\Gamma_\omega(q^2)}Ae^{i\phi}.\label{fbando} \\
\ee
The original HLS model preserved isospin symmetry and so did not
include the $\omega$. Isospin breaking has recently been studied in a
generalisation of the
HLS model \cite{hash}, however, here we have for simplicity employed
the same $\omega$ terms as  used for VMD.
The relations equivalent to Eqs.~(\ref{vmd2g}) and (\ref{epsy})
for the $\rho$ meson are now
\be
g_{\rho \gamma}=\frac{a}{2} \frac{m_{\rho}^2}{g_{\rho \pi
\pi}}=\frac{m_{\rho}^2}{g_\rho}\ .
\label{epsa}
\ee
We see that setting $a=2$ reproduces VMD2 in the limit of exact universality.
However, we wish to keep $a$ as a free parameter which we can fit to the data.
Note that in the HLS model universality violation and the existence of
a non--resonant coupling $\gamma \pi^+ \pi^-$ are related.
Note also that universality violation can be introduced
in the HLS model without violating the constraint
$F_\pi (0)=1$, in a natural way.

The muon form factor for the HLS model is exactly the same as for VMD2
(see Eqs.~(\ref{muonff}) and (\ref{epsa})).

\subsection{WCCWZ Lagrangian}
Birse has recently discussed \cite{birse} the pion form factor arising from
the WCCWZ Lagrangian \cite{WCCWZ} in which the vector and
axial vector fields transform homogeneously under non-linear chiral
symmetry. The scheme imposes no constraints on the couplings of the
spin 1 particles beyond those of approximate chiral symmetry. Birse's version
of the form factor is (isospin violation is not considered)
\be
F_\pi(q^2)=1-\frac{g_1f_1}{f_\pi^2}\frac{q^4}{q^2-m^2_\rho+im_\rho
\Gamma_\rho(q^2)}+\frac{f_2}{f_\pi^2}q^2,
\label{wccwzo}
\ee
where the first two terms on the RHS are those arising from the
WCCWZ Lagrangian.
The $q^4$ piece grows at large $q^2$ in a way incompatible with
QCD predictions (for a discussion of matching the asymptotic prediction
to a low energy model see Geshkenbein \cite{Gesh}).
 The $f_2$ contribution has been added by Birse to modify
this high energy behaviour toward that expected in QCD. To this end,
Birse sets
\be
f_2=g_1f_1=\frac{f_\pi^2}{m^2_\rho}\label{birsec}
\ee
and recovers the universality limit of the form factor
in which VMD1 and VMD2 are equivalent
(in the zero width approximation).  Note that a $q^2$-dependence of
the non-resonant background is what one would, in general,
obtain from the WCCWZ framework, implemented in its most general form,
which relies only on the symmetries
of QCD.  In constructing a phenomenological implementation we have, however,
simplified the most general form, adding what amounts to
a minimal $q^2$-dependence to the background term of VMD1.  The
resulting form factor, which we will refer to as the WCCWZ model, is then
\be
F_\pi(q^2)=1+b q^2-\frac{g_{\rho\gamma}^{\rm WCCWZ}
(q^2)g_{\rho\pi\pi}}{q^2-m^2_\rho+im_\rho
\Gamma_\rho(q^2)} - g_{\omega \gamma}
\frac{Ae^{i\phi}}{q^2-m_\omega^2+im_\omega\Gamma_\omega(q^2)},
\label{WZM1}
\ee
where we keep $b$ as an independent parameter to be fit.

The WCCWZ model, thus, has one more free parameter
than VMD1. We have added
the $\omega$ contribution as above for all other models. The muon
form factor is exactly the same as for VMD1 (see Eq.~(\ref{muonff})).

It is important to note that the WCCWZ model allows one
to have a non--resonant
term which can be mass dependent, cf., VMD2 which
carries only resonant contributions or VMD1 or the HLS models
which both exhibit only constant non--resonant contributions to the
pion form factor. The expression in Eq.~(\ref{WZM1}) exhibits an unphysical
high energy behaviour; however, as we are only interested in the
pion form factor at low energies (below 1 GeV), this feature is not relevant.
Of course, one can consider that such a polynomial structure at low
energies represents an approximation in the resonance region to
a function going to zero at high energies
{\footnote {For instance, $1+bq^2$ can be considered as
the first terms of the Taylor expansion of a function like $1/(1-bq^2)$
as suggested by \cite{chung}.}}.

\indent The model VMD2 contains only resonant contributions, whereas VMD1,
HLS and WCCWZ also contain a non--resonant part. For VMD1 and HLS, this
term is constant ({\it} i.e., pointlike) as in standard lowest order
QED. We shall frequently refer to this term as a direct $\gamma\pi\pi$
contribution or coupling.  In the case
of WCCWZ, this non--resonant term also contains a $q^2$--dependent piece
which clearly indicates a departure from a point--like coupling; nevertheless,
we shall also  refer to it as
a direct $\gamma\pi\pi$ coupling for convenience.

\subsection{Elastic Unitarity}
\label{unit}

When $\lambda\neq1$
the pion form factor described above in the VMD, HLS  and WCCWZ
models does not in general {\em exactly} fulfill unitarity.
This is because  one employs simultaneously
a bare, undressed $\rho^0\pi^+\pi^-$ coupling (given by the tree--level
coupling, $g_{\rho\pi\pi}$) and a dressed $\rho^0$ propagator, as
signalled by the presence of a momentum--dependent width in Eq.~(\ref{mesnwid})
(see, {\it e.g.}, Ref.~\cite{klingl}).  

One can then show\cite{benayoun} that unitarity can be restored
to the model treatments above by replacing
the coupling $g_{\rho \pi \pi}$ in Eqs.~(\ref{ff1}), (\ref{ff2}),
(\ref{fbando})
and (\ref{WZM1}) with
\be
G_{\rho}(q^2)=\displaystyle
\sqrt{6 \pi
 \frac{m_{\rho}q}{p_{\pi}^3(q^2)}
\Gamma_{ \rho \ra \pi^+ \pi^-}(q^2)}
\label{unit6}
\ee
where $q\equiv\sqrt{q^2}$.  This replacement leads to the
unitarised versions of our VMD models.
The connection between this ``dressed"
vertex function and $g_{\rho \pi \pi}$ gives (cf., Eq.~(4.16) of
Ref.~\cite{klingl})
\be
g_{\rho \pi \pi}=\sqrt{6 \pi
\frac{ m_{\rho}^2}{p_{\pi}^3(m_{\rho}^2)}\Gamma_{\rho}} ,
\label{vtx1}
\ee
from which we see that
\be
g_{\rho \pi \pi}=   G_{\rho}(m_{\rho}^2).
\label{vtx2}
\ee

It should be noted that the left hand side of Eq.~(\ref{unit6})
becomes constant -- and then coincides with $ g_{\rho \pi \pi}$ --
if and only if
$\lambda=1$ (see Eq.\ (\ref{mesnwid})). Therefore, if  $\lambda \equiv 1$
Eqs.~(\ref{ff1}), (\ref{ff2}), (\ref{fbando}) and (\ref{WZM1})
are already unitarised.

\subsection{Phase of $F_\pi(q^2)$
and Phase of the $\pi \pi \ra \pi \pi$ Amplitude}\label{formfc}

\indent \indent
{From} general properties of field theory (mainly, unitarity and $T$--invariance),
it can be shown \cite{gasio} that for $s \equiv q^2$ real above
threshold, we have

\be
F_{\pi}(s)=\exp{[2i\delta_1^1]}~~F_{\pi}^*(s)~~~,
\label{formfc1}
\ee

\noindent
up to the first open inelastic threshold.
Therefore the phase of $F_{\pi}(s)$ allows
us to  extract the exact behaviour of the $\pi \pi \ra \pi \pi$
phase ($\delta_1^1$) in the region just above threshold .

\section{Simultaneous Fits of $e^+e^- \ra \pi^+ \pi^-$ and $\mu^+ \mu^-$ Data}
\label{fitcond}

Fitting the $e^+e^-$ data from threshold to about 1 GeV involves
three well-known resonances, $\rho^0$, $\omega$ and $\phi$. As the last
two are narrow, their parametrisation
is relatively simple. But, due to its broadness, the $\rho^0$
meson has given rise to long standing problems of parametrisation (see
Refs.~\cite{benayoun,GS,PRoos} and previous references
quoted therein). Moreover, as we have mentioned,
one can ask whether experimental data require the existence of
a non--resonant $\gamma\pi^+ \pi^-$ coupling.
The conclusion of Refs.~\cite{benayoun,benayoun2} is that
 data on $F_\pi(q^2)$ alone are insufficient to answer this question.

In addition to the mass and width, in the context of the class
of models having widths of the form given in Eq.~(\ref{mesnwid}), one requires
 only
one additional parameter, $\lambda$, to define the $\rho$ shape.
The resulting
fit turns out to depend not only on the
non-resonant
coupling, but also on whether unitarisation is used or not.
One approach \cite{IS} to this problem is to
perform a simultaneous fit of all $e^+ e^- \ra \pi^+ \pi^-$ data \cite{barkov}
and $e^+ e^- \ra \mu^+ \mu^-$ data \cite{shwartz}. Indeed, if the data are
precise enough, we could see a non--resonant coupling  in $e^+ e^- \ra
\pi^+ \pi^-$, which will (of course) be small in
$e^+ e^- \ra \mu^+ \mu^-$.  Therefore, from first
principles, a simultaneous fit to both data
allows us to decouple the $\rho$ from  any non--resonant
$\gamma \pi^+ \pi^-$ coupling. Naturally, to be of any use in this,
the \emuon data would have to be very good.
Until recently relatively precise
measurements were available only for the region
around the $\phi$ mass \cite{orsay,kurdadze}. However, a new
data set \cite{shwartz} collected by the {\sc olya}
collaboration is available and covers a large invariant mass interval
from 0.65 GeV up to 1.4 GeV. We shall see shortly whether
it is precise enough to constrain the $\rho$ parameters.
Thus, the data sets which will be used for our fits are those
collected by {\sc dm1}, {\sc olya} and {\sc cmd} which are
tabulated in \cite{barkov} (for \ep) and only the {\sc olya} data
of \cite{shwartz} for \emuon; these  data sets do not cover the
$\phi$ peak region.

\section{Results of $e^+e^-$ data analysis}
\label{fitee}

In all of the previously described models, except for WCCWZ, the fit to
$e^+ e^- \ra \pi^+ \pi^-$ and $e^+ e^- \ra \mu^+ \mu^-$  data depends on
only five parameters.
The first four are the three $\rho$ meson parameters
(${m}_{\rho}$, $\Gamma_{\rho}$ and $\lambda$) and the
Orsay phase ($\phi$).
These are common to both VMD and HLS. In the VMD models
an additional parameter, $\epsilon$, has been introduced
in order to account for universality violation (see Eqs.~(\ref{epsy})
and (\ref{1vgam})),
while in the HLS model this parameter is replaced by $a$ (see
Eq.~(\ref{epsa})).
These last parameters allow us to fit the branching fraction
$\rho^0 \ra e^+ e^-$ within each model in a consistent way.
The WCCWZ model
depends on one additional parameter, $b$, which permits
a more flexible form for the non--resonant contribution, as
compared with the VMD1 or HLS models. Let us note that
introducing \emuon in our fit procedure together with \ep
does not require further free parameters.

Generally speaking, the parameter named $A$ in the VMD models above determines
the branching ratio Br$(\omega \ra \pi^+ \pi^-)$ and should be set free since
its value is strongly influenced by the data on \ep we are fitting. However,
in order to minimise the number of fit parameters at the stage when different
models are still considered, we fix its value from the corresponding world
average value \cite{PDG} of Br$(\omega \ra \pi^+ \pi^-)$.
We shall set $A$ free for our last fit, in order to get an optimum estimate of
Br$(\omega \ra \pi^+ \pi^-)$; this will be done only for the model
which survives all selection criteria.

Finally the fits have been performed for both the standard VMD, HLS
and WCCWZ models and their unitarised versions, for both
$e^+ e^- \ra \pi^+ \pi^-$ data alone and simultaneously with
$e^+ e^- \ra \mu^+ \mu^-$.
As all measurements in the region of the $\phi$ resonance
for each of these final states are not published
as cross sections \cite{orsay,kurdadze}, they are not taken into
account in our fits. When fitting the data, we take into account
the statistical errors given in \cite{barkov} for each \ep
data set. {\sc dm1} and {\sc cmd} claim negligible systematic errors
(2.2\% for {\sc dm1} and 2\% for {\sc cmd}, while the statistical errors
are typically 6\% or greater); these errors can thus
be neglected with respect to the quoted statistical errors. {\sc olya}
claims smaller statistical errors but larger systematic errors:
these two errors have comparable magnitudes from the $\rho^0$ peak
to the $\phi$ mass. We do not expect a dramatic influence from neglecting
these systematic errors, except that this would somewhat increase
the $\chi^2$ value at minimum and hence worsen slightly the fit quality.

The results are displayed in Table \ref{table1} (non--unitarised models) and
in Table \ref{table2} (unitarised models).
We show the fitted parameters in the upper section of each
table, while in the lower part we
provide the corresponding values for derived parameters of relevance.

\begin{table}[htb]
\begin{center}
\begin{tabular}{|c||c|c|c|c|}
\hline\hline
\hhhb {Parameter} & {VMD1 } & VMD2 & HLS & PDG  \\
\hline\hline
\hhhb
$\epsilon$ & $0.210^{+0.016}_{-0.018}$ & $0.163^{+0.007}_{-0.008}$ & --
& -- \\
\hhha
HLS $a$ & -- & -- & $2.399^{+0.028}_{-0.012}$ & --\\
\hhha
$m_{\rho}$ (MeV)      & $751.4^{+3.7}_{-2.8}$ & 776.74$\pm$2.2
&$755.1^{+4.9}_{-2.8}$ &  $769.1 \pm 0.9$ \\
\hhha
$\Gamma_{\rho}$ (MeV) &$146.0\pm 2.2$ & $145.10^{+2.1}_{-1.9}$ &
$143.32^{+1.8}_{-2.0}$ &151.0 $\pm$ 2.0\\
\hhha
$\phi$ (degrees)&$113.8^{+5.2}_{-6.9}$ &$106.3\pm 4.5$&$120.6^{+4.6}_{-5.7}$
&
-- \\
\hhha
$\lambda$ & $4.49^{+0.35}_{-0.43}$ & $1.61^{+0.34}_{-0.31}$ &
$3.92^{+0.34}_{-0.72}$& -- \\
\hhha
$\chi^2/{\rm dof}$ ($\pi \pi$) & 63/77 &148/77 &64/77 & -- \\
\hhha
$\chi^2/{\rm dof}$ ($\pi \pi + \mu \mu$) &105/115 &194/115 &108/115 & -- \\
\hline\hline
\hhha
$g_{\rho \gamma}$ (GeV$^{2}$) & $0.113 \pm 0.003$ & $0.119 \pm 0.002$ &
$0.115
\pm 0.003$ &0.120 $\pm$ 0.003\\
\hhha
$g_{\omega \gamma}$ (GeV$^{2}$)& -- & -- & -- & 0.036 $\pm$ 0.001\\
\hhha
$g_{\rho\pi\pi}^2/4\pi$ & $2.91 \pm 0.05$ & $2.76 \pm 0.03$  & $2.84
\pm 0.05$ &--\\
\hhha
$\Gamma(\rho \ra e^+e^-)$ (keV) & $6.72^{+0.38}_{-0.33}$& $6.74 \pm 0.20$ &
$6.34^{+0.45}_{-0.27}$& $6.77\pm 0.32$\\
\hline\hline
\end{tabular}
\parbox{130mm}{\caption{Results from fits to $F_\pi(q^2)$ and
$F_\mu(q^2)$ without unitarisation for the VMD1, VMD2,
and HLS models.  Shown for comparison are the Particle Data Group
quoted values \protect\cite{PDG}.
}
\label{table1}}
\end{center}
\end{table}

\begin{table}[htb]
\begin{center}
\begin{tabular}{|c||c|c|c|c|}
\hline\hline
\hhhb {Parameter} & {VMD1 } & VMD2 & HLS & WCCWZ  \\
\hline\hline
\hhhb
$\epsilon$ & $0.167 \pm 0.008$ & $0.215\pm 0.010$ & -- &
$0.142\pm 0.014$ \\
\hhha
HLS $a$ & -- & -- & $2.364 \pm 0.015$ & --\\
\hhha
WCCWZ $b$ (GeV$^{-2}$) & -- & -- & -- & $-0.319^{+0.139}_{-0.117}$\\
\hhha
$m_{\rho}$ (MeV)      & $774.67 \pm 0.65$ & $780.37 \pm 0.65 $
&$775.15 \pm 0.65$ &  $770.89^{+1.75}_{-1.51}$ \\
\hhha
$\Gamma_{\rho}$ (MeV) &$147.11\pm 1.60$ & $155.44 \pm 1.95$ & $147.67
\pm 1.47 $& $140.6^{+3.2}_{-2.9}$\\
\hhha
$\phi$ (degrees)&$94.7\pm 4.3$ &$98.8\pm 4.4$&$105.1 \pm 4.3$&
$101.7\pm 5.3$ \\
\hhha
$\lambda$ & $1.038^{+0.080}_{-0.085}$ & $0.567 \pm 0.055$
&$1.056 \pm 0.042$ & $1.623^{+0.231}_{-0.269}$ \\
\hhha
$\chi^2/{\rm dof}$ ($\pi \pi$) & 65/77 &81/77 &65/77 & 61/76 \\
\hhha
$\chi^2/{\rm dof}$ ($\pi \pi + \mu \mu$) &104/115 &128/115 &111/115 &
103/114\\
\hline\hline
\hhha
$g_{\rho \gamma}$ (GeV$^{2}$) & $0.118 \pm 0.001$ & $0.122 \pm 0.001$ &
$0.114\pm 0.001$ & $0.133\pm 0.007$\\
\hhha
$g_{\rho\pi\pi}^2/4\pi$ & $2.81 \pm 0.03$ & $2.94 \pm 0.04$  & $3.08
\pm 0.03$ & $2.08\pm 0.04$\\
\hhha
$\Gamma(\rho \ra e^+e^-)$ (keV) & $6.70 \pm 0.11$& $6.99 \pm 0.16$ &
$6.23 \pm 0.11$& $8.62\pm 0.46$\\
\hline\hline
\end{tabular}
\parbox{130mm}{\caption{Results from fits to \protect{$F_\pi(q^2)$}  and
\protect{$F_\mu(q^2)$} for the unitarised VMD1, VMD2, HLS and
WCCWZ models.}
\label{table2}}
\end{center}
\end{table}

We find that, disappointingly, the new muon data\cite{shwartz} places
no practical constraint on the $\rho$ parameters extracted from $F_\pi(q^2)$.
Another striking conclusion is that it is generally possible to achieve
a very good fit to the pion data whichever model is used, unitarised
or not; the single exception to this being non-unitarised VMD2
(which is the usual model for $\rho$ physics).  Correspondingly,
the significant model dependence of the extracted $\rho$ mass should be
noted.

As far as one relies only on the statistical quality of fits  for the cross
section $e^+ e^- \ra \pi^+ \pi^-$, the existing data do not allow one
to determine the most suitable way to implement vector meson dominance,
except  to discard the non--unitarised version of VMD2 which is
clearly  disfavoured by the data.
On the other hand, the possible values for the $\rho^0$ mass
cover a wide mass range: from 750 MeV to 780 MeV.
The single firm conclusion which can be drawn from the above
analysis is that, whichever is the VDM parametrisation
chosen, unitarised or not, one always observes a small but statistically
significant signal of universality violation: $\epsilon \simeq 0.20$
(instead of 0) or $a \simeq 2.4$ (instead of 2).

The question, therefore, remains as to whether it is possible to find
other criteria to distinguish between the various ways of building
effective Lagrangians involving the vector mesons.

\section {Comparison with $\pi \pi$ Phase shift Analysis}
\label{phaseshift}

It is of some interest to compare the phase shifts
predicted from  the versions of VMD obtained after fitting to the
$e^+ e^- \ra \pi^+ \pi^-$ data  to the values of $\delta^1_1$
as tabulated by Ref.~\cite{petersen} (see their Table 1).
In this way, it is possible to check the consistency
of the information deduced assuming each VMD formulation
with the results of Ref. \cite{petersen} which were derived under
completely independent assumptions (namely, unitarity, analyticity
and crossing symmetry). 

\begin{figure}[ht]
  \centering{\
     \epsfig{angle=0,figure=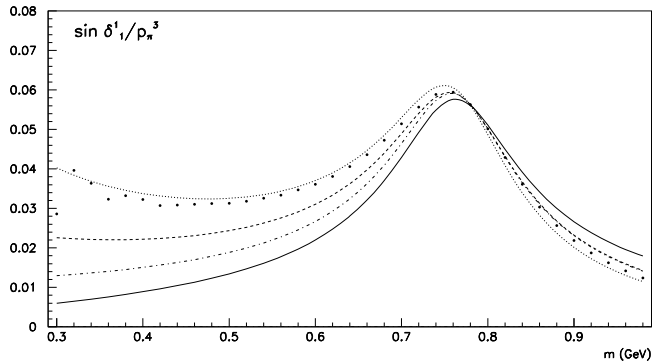,width=0.8\linewidth}
               }
\parbox{130mm}{\caption
{Function $\sin{\delta^1_1}/p_{\pi}^3$ deduced from $\pi^+ \pi^-$
phase shifts; the function is given in units of $m_{\pi}^{-3}$.
The dots correspond
to the points of Ref.\ \protect\cite{petersen}. Full line is VMD1,
dashed line is VMD2, dotted line is HLS (the best description
of expected phase) and dashed--dotted is WCCWZ.}
\label{figadel5}  }
\end{figure}

A sensitive way to carry on the comparison of the $F_{\pi}(q^2)$ phases
with the Froggatt--Petersen phase shift, is to
superimpose the  various model predictions for
$\sin{\delta^1_1}/p_{\pi}^3$.
This function is connected with the  isospin 1 $P$--wave
scattering length $a_1^1$ through:
\begin{equation}
a_1^1= \displaystyle \lim_{q^2 \ra 4 m_{\pi}^2}
\frac{\sin{\delta^1_1(q^2)}}{p_{\pi}^3(q^2)}
\label{scatlen1}
\end{equation}
and  has the  characteristic of highly magnifying differences due
to the phases in the low energy region (up to, say, 600 MeV).
The curves corresponding
to the various unitarised models are shown in Fig.~\ref{figadel5}.

\section{Summary and Conclusions}
\label{conclusion}

We have studied a
 variety of vector meson dominance models
in both non--unitarised and unitarised forms. They depend on a few
parameters (mass of the $\rho$ meson, its coupling constants to
$\pi \pi$ and $e^+ e^-$, the shape parameter $\lambda$ and the Orsay phase
needed in order to describe the $\rho-\omega$ mixing). We have fitted these
to both $e^+e^-\rightarrow \pi^+\pi^-$ and $e^+e^-\rightarrow \mu^+\mu^-$.
In order to study the behaviour of each solution, we have studied
how they match the $\pi \pi$ phase shift obtained under general
model independent assumptions from threshold up to 1 GeV. We have
also examined \cite{benayounetal}
the value they provide for threshold parameters
($F_{\pi}$ at threshold and the scattering length $a^1_1$),
which can be estimated accurately from ChPT.

This represents the largest set of independent data
and cross--checks done so far. It happens that,
of the models considered, only the
unitarised HLS model is able to account for all examined
effects. We also find that the standard value
$\lambda=1$, corresponding to a point-like $\rho\pi\pi$ coupling,
is well accepted by the data for $\rho$ parametrisation.

Unlike the standard formulation of VMD, fits with this model
return a significant
non--resonant
contribution to the electromagnetic pion form factor.
This was found to have a value
$\simeq -e/6$.  This term is governed
by a small universality violation which changes
the HLS parameter $a$ from 2 to 2.4. All other models considered,
even if { they are} able to describe
$e^+e^-$ annihilations quite well,
are unable to account satisfactorily for the other available information.
It should be mentioned that, within the class of models considered,
our results tend to favor a constant
$g_{\rho \gamma}$ over a $q^2$ dependent one.

We conclude that, of the models considered, the HLS model with
$a \simeq 2.4$ is the most favoured version for
implementing the VMD ansatz. Thus, it is interesting to consider
whether the success of its predictions for the magnitude
and phase of $F_\pi(q^2)$, could be
obtained in a way which does not need the assumption that the $\rho$ is
a dynamical gauge boson of a hidden local gauge symmetry.
Hence, looking for other models able to describe,
as successfully as the HLS model, the same large set of data is
useful in order to know whether the conceptual
motivation for this model { should be interpreted as having any
underlying significance.}

\section*{Acknowledgments} 
I would like to acknowledge and thank my collaborators on the work
reported here, i.e., M.\ Benayoun, S.\ Eidelman, K.\ Maltman,
H.B.\ O'Connell, and B.\ Schwartz.  I would also like to thank the
APCTP for its hospitality during the workshop and to offer Mannque
Rho my warm congratulations on this celebration of his 60$^{\rm th}$ birthday. 
This work was supported by the Australian Research Council.

\end{document}